\documentclass{article}

\usepackage{arxiv}

\usepackage[utf8]{inputenc} 
\usepackage[T1]{fontenc}    
\usepackage{xcolor}
\usepackage{hyperref}       
\usepackage{url}            
\usepackage{booktabs}       
\usepackage{amsfonts}       
\usepackage{nicefrac}       
\usepackage{microtype}      
\usepackage{graphicx}
\usepackage{natbib}
\usepackage{doi}
\usepackage{wrapfig}
\usepackage{multicol}

\hypersetup{
  colorlinks,
  citecolor=blue,
  linkcolor=blue,
  urlcolor=blue}

\title{Robustness of AI-based weather forecasts in a changing climate}

\date{September 27, 2024}	
\author{\href{https://orcid.org/0000-0002-5468-575X}{\includegraphics[scale=0.06]{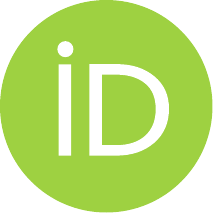}\hspace{1mm}Thomas Rackow} \\
	European Centre for Medium-Range\\ Weather Forecasts (ECMWF)\\
	\texttt{thomas.rackow@ecmwf.int}
	\And
	\href{https://orcid.org/0000-0002-3365-8146}{\includegraphics[scale=0.06]{orcid.pdf}\hspace{1mm} Nikolay Koldunov} \\
	Alfred Wegener Institute\\
	  Helmholtz Centre for Polar\\and Marine Research (AWI)\\
	\texttt{nikolay.koldunov@awi.de}
    \And
 	\href{https://orcid.org/0000-0002-2740-6815}{\includegraphics[scale=0.06]{orcid.pdf}\hspace{1mm} Christian Lessig} \\
	European Centre for Medium-Range\\ Weather Forecasts (ECMWF)\\
	\texttt{christian.lessig@ecmwf.int}
 	\And
	\href{https://orcid.org/0000-0002-1215-3288}{\includegraphics[scale=0.06]{orcid.pdf}\hspace{1mm} Irina Sandu} \\
	European Centre for Medium-Range\\ Weather Forecasts (ECMWF)\\
     \And
        \href{https://scholar.google.it/citations?user=gk-SRkgAAAAJ}{\hspace{1mm} Mihai Alexe} \\
       European Centre for Medium-Range\\ Weather Forecasts (ECMWF)\\
    \And
      	\href{https://orcid.org/0000-0002-1132-0961}{\includegraphics[scale=0.06]{orcid.pdf}\hspace{1mm} Matthew Chantry} \\
	European Centre for Medium-Range\\ Weather Forecasts (ECMWF)\\
   	\And
      	\href{https://orcid.org/0000-0002-5010-0363}
       {\includegraphics[scale=0.06]{orcid.pdf}\hspace{1mm} Mariana Clare} \\
	European Centre for Medium-Range\\ Weather Forecasts (ECMWF)\\
   	\And
      	\href{https://orcid.org/0000-0001-8273-905X}{\includegraphics[scale=0.06]{orcid.pdf}\hspace{1mm} Jesper Dramsch} \\
	European Centre for Medium-Range\\ Weather Forecasts (ECMWF)\\
  	\And
	\href{https://orcid.org/0000-0003-1766-2898}{\includegraphics[scale=0.06]{orcid.pdf}\hspace{1mm} Florian Pappenberger} \\
	European Centre for Medium-Range\\ Weather Forecasts (ECMWF)\\
  	\And
   	\href{https://orcid.org/0000-0001-5129-6364}{\includegraphics[scale=0.06]{orcid.pdf}\hspace{1mm} Xabier Pedruzo-Bagazgoitia} \\
	European Centre for Medium-Range\\ Weather Forecasts (ECMWF)\\
  	\And
	\href{https://orcid.org/0000-0002-2961-0289}{\includegraphics[scale=0.06]{orcid.pdf}\hspace{1mm} Steffen Tietsche} \\
	European Centre for Medium-Range\\ Weather Forecasts (ECMWF)\\
  	\And
	\href{https://orcid.org/0000-0002-2651-1293}{\includegraphics[scale=0.06]{orcid.pdf}\hspace{1mm} Thomas Jung} \\
    Alfred Wegener Institute (AWI)\\
    University of Bremen
}

\date{}

\renewcommand{\shorttitle}{Robustness of AI forecasts in a changing climate}

\hypersetup{
pdftitle={AI forecasts in a changing climate},
pdfsubject={manuscript},
pdfauthor={T. Rackow et al.},
pdfkeywords={numerical weather prediction, climate projection, machine learning, deep learning, artificial intelligence},
}

\begin{document}
\maketitle


\begin{abstract}
    Data-driven machine learning models for weather forecasting have made transformational progress in the last 1--2 years, with state-of-the-art ones now outperforming the best physics-based models for a wide range of skill scores.
    Given the strong links between weather and climate modelling, this raises the question whether machine learning models could also revolutionize climate science, for example by informing mitigation and adaptation to climate change or to generate larger ensembles for more robust uncertainty estimates.
    Here, we show that current state-of-the-art machine learning models trained for weather forecasting in present-day climate produce skillful forecasts across different climate states corresponding to pre-industrial, present-day, and future 2.9K warmer climates. This indicates that the dynamics shaping the weather on short timescales may not differ fundamentally in a changing climate.
    It also demonstrates out-of-distribution generalization capabilities of the machine learning models that are a critical prerequisite for climate applications.
    Nonetheless, two of the models show a global-mean cold bias in the forecasts for the future warmer climate state,\,i.e. they drift towards the colder present-day climate they have been trained for. 
    A similar result is obtained for the pre-industrial case where two out of three models show a warming. We discuss possible remedies for these biases and analyze their spatial distribution, revealing complex warming and cooling patterns that are partly related to missing ocean-sea ice and land surface information in the training data. 
    Despite these current limitations, our results suggest that data-driven machine learning models will provide powerful tools for climate science and transform established approaches by complementing conventional physics-based models.
\end{abstract}


\section*{Main}

\begin{figure}
  \centering
\includegraphics[width=\textwidth]{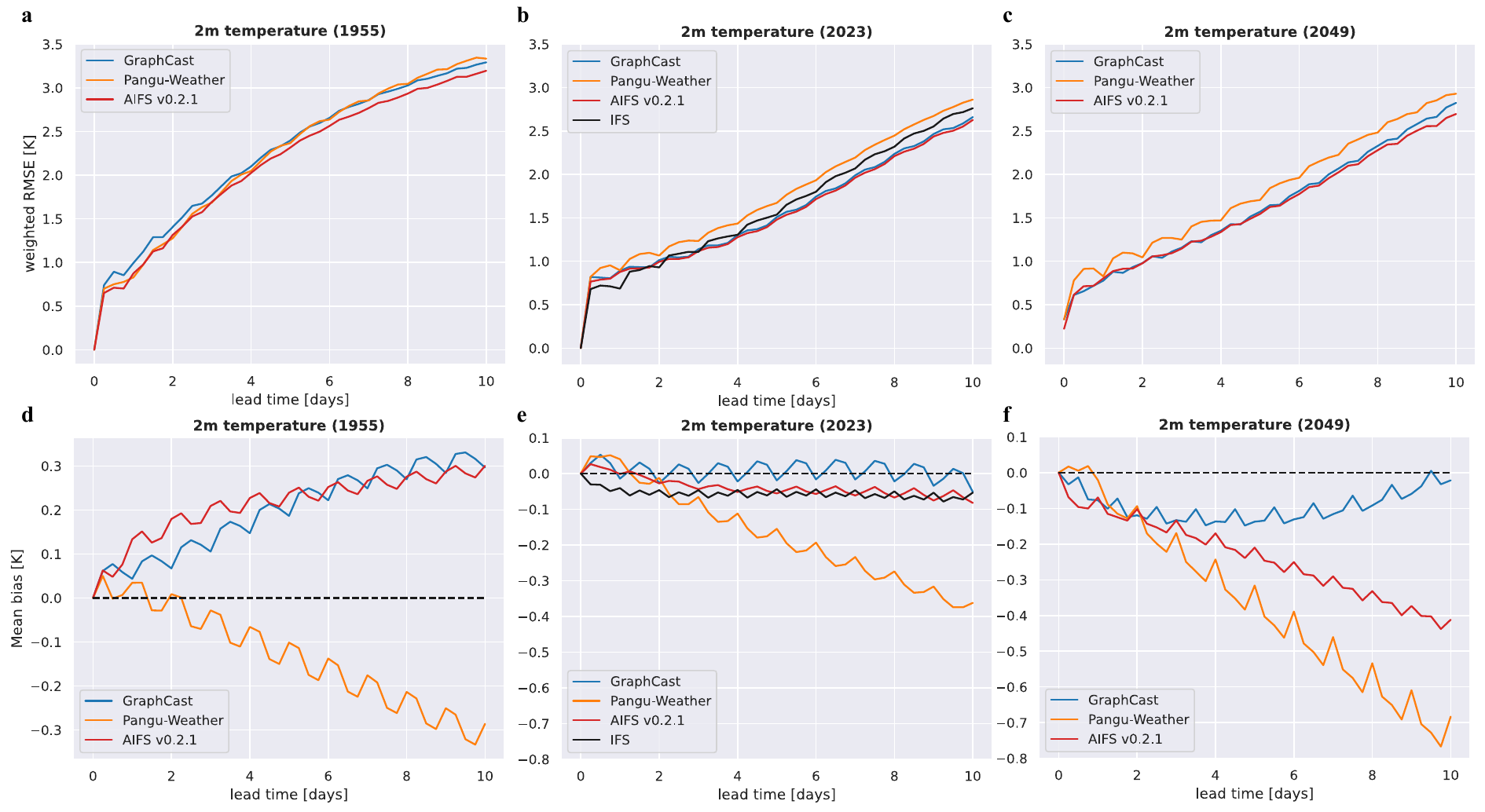}
  \caption{\textbf{Skill of data-driven 10-day weather forecasts for 2m temperature in conditions representing pre-industrial, present-day, and future +2.9K climate states, for AIFS, GraphCast, and Pangu-Weather.} Root-mean square error (RMSE, in K) for all models, weighted by the cosine of latitude in \textbf{a}, 1955, a proxy-year for pre-industrial times, \textbf{b}, in 2023, representing present-day climate, and \textbf{c}, in 2049, the +2.9K warmer world. \textbf{d}, Mean bias [K] for all models in 1955, \textbf{e}, in 2023, and \textbf{f}, in 2049. In 1955, Pangu-Weather cools further while the other two models show a warming. 
  In 2023, GraphCast and AIFS show very little bias, similar to ECMWF's operational IFS forecast after 10 days, while Pangu-Weather consistently cools. In 2049, positive and negative biases for GraphCast balance while the other two models are characterised by a cold bias.
  All RMSE and bias evolutions in 1955, 2023, and 2049 are averaged over the daily forecasts (see Methods).}
  \label{fig:rmse}
\end{figure}

Climate projections based on Earth system models are critical to inform assessment reports by the Intergovernmental Panel on Climate Change (IPCC) \citep{lee2023ipcc} and they drive policies for climate change adaptation and mitigation. Yet, the corresponding simulations remain computationally expensive~\citep{Bauer2021aa} and suffer from persistent biases \citep{NorthwestCorner2015,Milinski2016,Rackow2019,Tian2020persistent,WilliamsScaife2023}. 
The last two years have seen a breakthrough in data-driven approaches for weather forecasting with machine learning models exploiting the Copernicus Climate Change Service's ERA5 reanalysis \citep{hersbach2020era5}, produced by ECMWF, as training dataset. These models are now being on par or even surpassing operational, physics-based models for a wide range of metrics~\citep{BenBouallegue2023rise,bi2023pangu,lam2023graphcast,lang2024aifs}. The data-driven models even outperform existing ones for predicting extreme events,
e.g. tropical cyclones~\citep{lam2023graphcast,lang2024aifs}, and have learned realistic Rossby wave dynamics \citep{Hakim2023}, a key ingredient for mean climate and its variability at mid-latitudes \citep{blackmon1976}.
At the same time, data-driven models require 3--4 orders of magnitude less computing time and energy to produce forecasts, once trained. Promising data-driven approaches towards seasonal prediction have emerged as well~\citep{Weyn2021,Chen:2024aa,Vonich2024}. Whether a similar success with data-driven models can be achieved for climate prediction remains, however, an open question. First promising results with stable control runs and forced atmospheric simulations with prescribed historical sea surface temperatures have been demonstrated~\citep{Wattmeyer2023ace,Kochkov2024neuralgcm}. 

For state-of-the-art data-driven weather forecasting models, 
one cannot assume a priori that they have learnt robust generalisable physics applicable to other climate states \citep{bonavita2023limitations}, 
as the models have been trained solely to minimise global-mean scores for short- or medium-range forecasting.
Nonetheless, individual studies have demonstrated remarkable generalization capabilities,\,e.g. providing skillful predictions for extreme events that are rare in the training data~\citep{bi2023pangu,lam2023graphcast,lang2024aifs}, producing physically consistent forecasts from idealized initial conditions~\citep{Hakim2023}, and performing downscaling from low-resolution climate model data without being trained for it~\citep{Koldunov2024}. 
The limits of the generalization abilities, including to different climate states, still need to be explored. This is of great importance in the context of climate projections where one wants to use the models for climate states outside of the 1979--2018 ERA5 training data with different external forcing. 

Skillful machine learning-based weather forecasting in different climate states would open the door for a wide range of climate applications. It can be used to produce a higher temporal resolution~\citep{Koldunov2024} for existing climate simulation output. Robust AI-based weather forecasting could also be employed in attribution studies on the influence of anthropogenic climate change where physics-based weather forecasting models have already shown promise, such as for heatwaves \citep{leach2024heatwave}. Running data-driven weather forecasts with initial conditions from a climate run, e.g.\,every day in a season for multiple consecutive years, creates an ensemble of potential weather states and hence allows one to derive weather statistics in support of uncertainty quantification. 
This includes information most pertinent for adaption and mitigation efforts to climate change, such as the change in intensity and impact maps of tropical cyclones, and the change in frequency of strong baroclinic systems. 
Robust weather forecasts in different climate states are also an important stepping stone for the development of machine learning-based climate projections. First steps for these have already been made~\citep{Wattmeyer2023ace,Guan2024,Wang2024coupled} but work such as the one we undertake in this study is required to ensure the reliability of these under changing external forcing. 

In this study, we explore three state-of-the-art data-driven weather forecasting models and their robustness to initial conditions from different climate states. For this, we use the models \emph{out-of-the-box} and perform medium-range (10-day) forecasts with initial conditions representing pre-industrial, present-day, and a future 2.9K warmer climate (compared to pre-industrial).
To our knowledge, this is the first time that AI-based weather forecasting models are shown to work well across climate regimes, with some current limitations that will be discussed in detail below.
By analysing three data-driven models and samples from three different climate states, we provide insights that are not limited to the specifics of the machine learning models or climate regimes.

\section*{Data-driven weather forecasts in colder and warmer climate conditions}

\subsection*{Experimental Setup}

For our experiments, we use ECMWF's data-driven forecasting model, the AIFS (Artificial Intelligence Integrated Forecasting System) in version 0.2.1 \citep{lang2024aifs}. It has been trained on the ERA5 reanalysis \citep{hersbach2020era5} for the years 1979-2018 and was fine-tuned on ECMWF's operational analysis for the years 2019 and 2020 (see Methods). 
We also employ two data-driven models developed by leading technology companies: Google DeepMind's GraphCast \citep{lam2023graphcast} and Huawei's Pangu-Weather \citep{bi2023pangu}. Both models were trained with ERA5 data from 1979–2017. GraphCast is fine-tuned on ECMWF's operational 9km forecasts from 2016 to 2021. 
All three data-driven models provide data at approximately 0.25$^{\circ}$ resolution and have shown present-day forecasting skill, for a set of scores, that is similar or even better than that of ECMWF's Integrated Forecasting System (IFS), which is considered as physics-based reference model in the field.
Moreover, it is known that AIFS and GraphCast have a very small mean bias in current present-day climate~\citep{Husain2024_ECCC} (see Methods for more details).

Present-day initial conditions for the year 2023 are taken from ECMWF's operational analysis, which was close to 1.5K warmer than pre-industrial levels \citep{Copernicus2023}.
In order to have another (colder) reference that is relatively well constrained by data, year 1955 in the back extension of the ERA5 reanalysis produced by the Copernicus Climate Change Service (C3S) at ECMWF \citep{Bell2021} serves as a proxy for pre-industrial climate ($\approx 1.4$\,K colder than 2023, see Methods). The future state is from year 2049 of a high-resolution, kilometre-scale IFS-FESOM scenario simulation for 2020--2049 performed in the European H2020 project nextGEMS (see Methods). The IFS-FESOM model has a spatial resolution of 9\,km in the atmosphere and approximately 5\,km in the ocean \citep{rackow2024}. It shows a globally $\approx 2.9$\,K warmer world compared to pre-industrial levels~\citep{Cycle4:2024} and is $\approx 1.5$\,K warmer compared to 2023. 

To examine the out-of-distribution generalization capabilities of the machine learning models for climate applications, 10-day weather forecasts were produced, by initializing at 12:00 UTC for every day in the three chosen years (1955, 2023, and 2049). 
Forecasts were evaluated against the datasets used as initial conditions, a common approach in numerical weather prediction.
Technical differences between the reference datasets (operational analysis, ERA5, and free-running IFS) are not a leading-order effect in the results documented below (see Methods).
We focus on the evolution of global 2-metre (2m) temperature, which is a standard diagnostic and the headline score for climate simulations.
As metrics for determining skill, global root-mean square error (RMSE) and mean bias are employed (see Methods).
The latter is standard for climate models and has also been used previously for the analysis of data-driven forecasting models.
We also analyse the spatial distribution of the biases, which provides important insights into process representations by the models.

\subsection*{Forecast skill in different climates: Analysis of global RMSE and mean bias}

\begin{figure} 
\centering
\includegraphics[width=\linewidth,trim={0 0.92cm 0 0},clip]{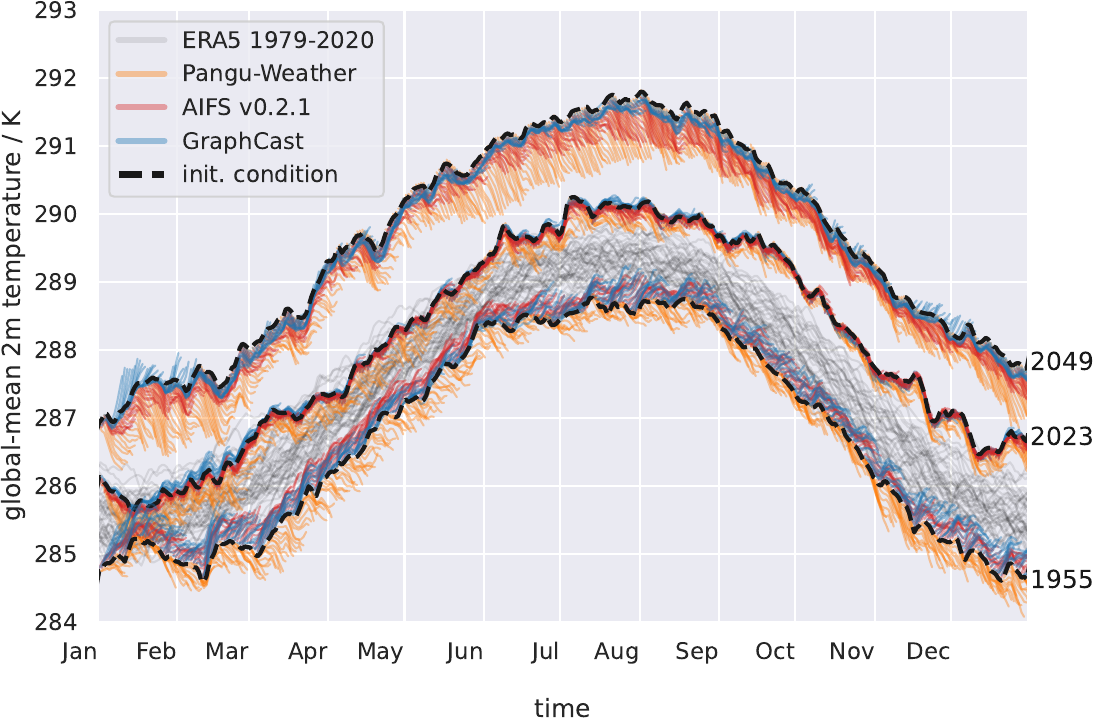}
\caption{\textbf{Global-mean 2m temperature evolution from daily 10-day weather forecasts under conditions representing pre-industrial (1955 as proxy), present-day (2023), and future +2.9K climate states (2049), for AIFS, GraphCast, and Pangu-Weather.} All models were trained with ERA5 reanalysis from 1979-2017 (with eventual fine-tuning), and the global-mean temperature for 1979--2020 is shown (light grey lines) for context. Black dashed lines are global-mean temperatures for 1955 (ERA5),  2023 (operational analysis), and for 2049 (physics-based scenario simulation), representing initial conditions for the data-driven forecast models. For every day in the three chosen years, 10-day forecasts were performed with every model starting from 12:00 UTC, resulting in 9 forecast datasets. The global-mean temperature of these 9 forecast datasets are shown as thin (hair-like) colored lines. 
For 1955, a warming of $0.3\,\mathrm{K}$ is found over 10 days for GraphCast and AIFS, while Pangu-Weather cools down. For 2023, none of the models shows any clear systematic bias. In 2049, AIFS and in particular Pangu-Weather cool substantially towards the climate conditions of the training period, at a rate of $-0.04\,\mathrm{K/day}$ and $-0.07\,\mathrm{K/day}$, respectively. Interestingly, GraphCast follows the climate projection in the global mean closely, but this is achieved by a compensation of cooling over land with a warming over the ocean.}
\label{fig:hairy}
\end{figure}
In general, the data-driven models produce skillful forecasts in the three different climates (Fig.\,\ref{fig:rmse}), despite the models being trained (or fine-tuned) on data from the period 1979-2021.
For 2023, the recent record-warm year \citep{Goessling2024recent}, AIFS, GraphCast, and Pangu-Weather all provide skillful forecasts comparable to the ECMWF operational IFS forecasts (Fig.\,\ref{fig:rmse}b). 
They also show no clear systematic biases in global-mean temperature, similar to the IFS forecasts after 10 days (Fig.\,\ref{fig:rmse}e), and produce forecasts close to the analysis throughout the year (Fig.~\ref{fig:hairy}). Only Pangu-Weather is too cold globally (Fig.\,\ref{fig:rmse}e), with a cooling of about -0.03\,K/day, consistent with previous findings \citep{BenBouallegue2023rise}. 
For the colder climate state represented by year 1955 that is prior to the training period, 2m temperature RMSE evolves similarly for all models (Fig.\,\ref{fig:rmse}a). While the RMSE evolution is inferior to the RMSE evolution in 2023, the models still provide skillful weather forecasts also in the 1.4\,K colder climate state. 
A warming of 0.3\,K can be observed for GraphCast and AIFS over the 10-day lead time (Fig.~\ref{fig:hairy} and Fig.\,\ref{fig:rmse}d), bringing the forecasts closer to the ERA5 data the models were trained on. 
Pangu-Weather again shows a global-mean cold bias (Fig.\,\ref{fig:rmse}d), cooling at an identical rate of -0.03\,K/day as seen for present-day conditions.
For the approximately 2.9\,K warmer climate in 2049, RMSE of 2m temperature over time is comparable to the RMSE evolution in 2023, which shows that skillful data-driven weather forecasts are possible also under future conditions (see Fig.\,\ref{fig:rmse}c). 
We hypothesize that the even lower RMSE for AIFS for the first two forecast days initialised from the 2049 data (compare Fig.\,\ref{fig:rmse}b,c) results from the scenario runs being based on IFS, and AIFS and GraphCast both being fine-tuned on  IFS data. 
This might enhance GraphCast's and AIFS's performance as an IFS-emulator.
For the warmer climate state, AIFS and in particular Pangu-Weather show a substantial cooling (compared to the physics-based model they were initialised from) towards the colder state in the training data (Fig.~\ref{fig:hairy}). In terms of mean bias, the cooling amounts to a rate of -0.04\,K/day and -0.07\,K/day, respectively (Fig.\,\ref{fig:rmse}f). 
Interestingly, GraphCast remains neutral in the global mean for the warmer climate state (Fig.~\ref{fig:hairy} and Fig.\,\ref{fig:rmse}f), although it results from compensating biases, as we discuss in the next section.

\subsection*{Drifting towards known climate state: spatial distribution of bias}

\begin{figure}
  \centering
\includegraphics[width=\textwidth]{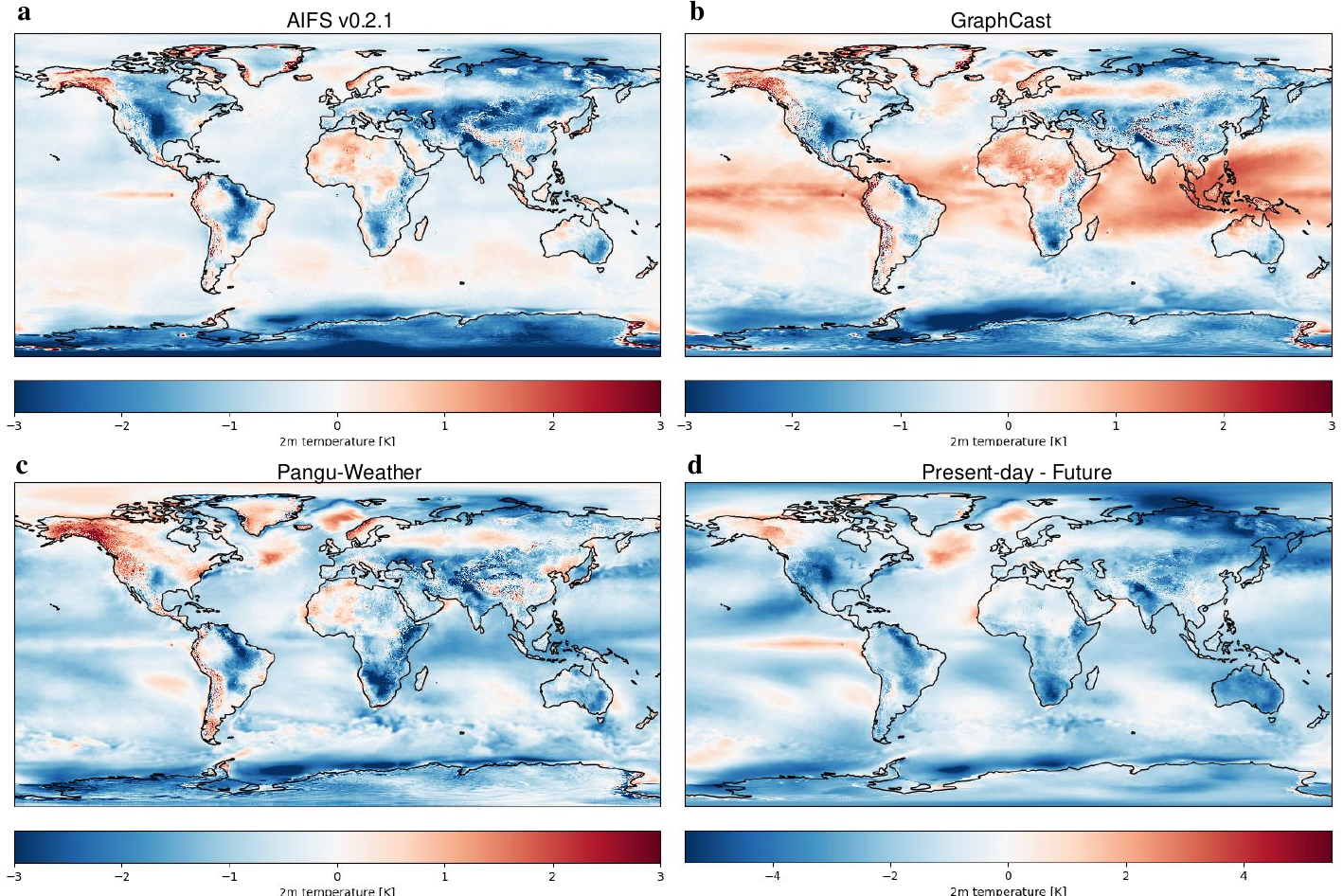}
\caption{\textbf{Mean 2m temperature drift over 10 days in 2049 from the data-driven models when compared to the reference physics-based simulation.} \textbf{a}, for AIFS v0.2.1, \textbf{b}, GraphCast, and \textbf{c}, for Pangu-Weather. \textbf{d}, Global 2m temperature difference between the present-day and future year used in this study as references (2023-2049, operational analysis minus scenario). Annual-mean fields for the data-driven models were constructed by combining 365 daily means from the end of their individual 10-day forecasts. The annual-mean from the reference simulation was then subtracted. 
There are many coherent areas with the same color in panels \textbf{a-c} and panel \textbf{d} where the models' drift towards the present-day conditions they were trained for aligns with the greatest differences in 2m temperature between future and present-day conditions.  
Note the different color range in panel \textbf{d} compared to the other panels.}
  \label{fig:maps-2049}
\end{figure}

To identify key areas of small or strong drift and to potentially obtain insights into the process representation of the data-driven weather forecasting models, we consider the spatial distribution of the bias (Fig.~\ref{fig:maps-2049} and Fig.~\ref{fig:maps1955-2023}).
For 2023, AIFS and GraphCast stay globally close to the analysis for a 10-day forecast (Supplementary Fig.\,\ref{fig:maps1955-2023}b,c) with 2m biases that on average resemble those of the operational IFS forecast at ECMWF (Supplementary Fig.\,\ref{fig:maps1955-2023}a). 
An exception are the polar regions, where GraphCast has a warm bias in the Arctic, and AIFS has a cold bias in polar regions where sea ice reached a low extent in 2023 and a record-low extent around Antarctica \citep{Roach2024seaice}.
Except over Antarctica and North America where biases are reaching 2--3\,K, Pangu-Weather shows very similar biases to the two other models over land, but cools consistently over the ocean (Supplementary Fig.\,\ref{fig:maps1955-2023}f).

For the proxy-preindustrial year 1955, AIFS and GraphCast in general warm over the ocean regions and on average also over land, in particular over the African Sahara, the North and South American West coasts, the Arabian Peninsula, and large parts of Asia (Supplementary Fig.\,\ref{fig:maps1955-2023}c,e). In contrast, Pangu-Weather cools  similarly to the behaviour it had shown in present-day climate with cooling concentrated over the ocean and no warming over the Sahara or Arabian Peninsula, in contrast to the other models.

When initialised from the warmer climate state in 2049 of the scenario simulation, all data-driven models had shown a global-mean 2m temperature cooling over a lead time of 10\,days, with Pangu-Weather cooling fastest (-0.07\,K/day), AIFS cooling second-fastest (-0.04 K/day), and GraphCast staying nearly neutral (Fig.\,\ref{fig:rmse}b).
Considering the spatial pattern of the 10-day changes, it can be seen that the neutral global-mean temperature anomaly by GraphCast is only achieved by compensating a cooling over land (with pattern very similar to AIFS and Pangu-Weather) with strong warming over the tropical oceans (Fig.\,\ref{fig:maps-2049}b). AIFS does not show this warming over the ocean and performs favourably there (Fig.\,\ref{fig:maps-2049}a). The general systematic cooling tendency of Pangu-Weather over the ocean irrespective of the initial state appears to add to the fact that Pangu-Weather cools at a much faster rate than the AIFS under future conditions (Fig.\,\ref{fig:maps-2049}a,c).
All data-driven models show a strong 10-day cooling over the Weddell Sea in the Southern Ocean, where the physics-based model warms strongly in 2049 compared to present-day conditions.
This is largely because of sea ice loss in this area (not shown) that cannot be modeled by the atmosphere-only data-driven weather forecasting models that drift towards known conditions of 2m temperature over a climatologically more extensive sea ice cover.

\subsection*{Relating future 2m temperature drift to global warming pattern}

To put the 2m cooling tendencies of the data-driven models applied in future conditions into perspective, we compute global-mean 2m temperature change between present-day conditions (from year 2023) and the future year 2049 used in this study (Fig.\,\ref{fig:maps-2049}d). In 2023, the data-driven models had shown best performance in terms of mean bias. Naturally, the variations between these single years are also influenced by specific weather events and internal variability. Therefore, our focus is on the patterns and signs, rather than the magnitudes or numerical values. The strong global-mean change of +1.5\,K between 2049 and 2023 allows to determine an estimate for the climate change pattern.

Notably, the 10-day 2m temperature drift over land in the data-driven models closely resembles the pattern of climate anomalies between the present-day and the future case (associated with both interannual variability and climate change), in particular the dipole over North America and South America, the pattern over Southern Africa, Ethiopia and Kenya, and the horse-shoe like pattern over large parts of Asia (compare Fig.\,\ref{fig:maps-2049}a,b,c to Fig.\,\ref{fig:maps-2049}d). This suggests that drift towards present-day conditions from the training dataset may be strongest where the differences in 2m temperature between future
and present-day conditions are largest. Over the ocean, the strong 2m temperature drift of Pangu-Weather resembles the climate anomalies very closely. With the exception of cooling over the Arctic and the Southern Ocean Weddell Sea, where sea ice is diminishing under global warming in the physics-based simulation, only the AIFS remains relatively unaffected by the significantly warmer ocean conditions in the future scenario, staying close to the physics-based simulation.

\section*{Discussion}

In this work, we address the question whether deep learning weather forecasting models trained on present-day data can provide skillful forecasts also in different colder and warmer states of the climate system. 
Answering this question would open the door for the use of AI-based forecasts in climate science. It would also lend increased confidence to machine learning models trained with historical data, allowing their application to various climate-related use cases in a rapidly warming world, from attribution to mitigation and adaptation studies in support of decision-making.
By using high-resolution scenario simulations as well as ERA5 reanalysis data and ECWMF's operational analysis as initial conditions, we show that three state-of-the-art data-driven weather forecasting models are indeed overall robust to the different initial conditions in terms of forecast scores, but more work needs to be done to further improve biases for climate applications.
With respect to global mean RMSE, very good forecasting skill is observed with AIFS, GraphCast, and Pangu-Weather for the recent past in 1955, for 2023, and in the middle of the 21st century in a potential climate that is approximately $1.5\,\mathrm{K}$ warmer than 2023 and about $2.9\,\mathrm{K}$ warmer than pre-industrial levels.
The skill of the data-driven machine learning models in different climates is astonishing, suggesting that there is no substantial change in the dynamics (e.g. of Rossby waves) driving the atmosphere on weather time scales in a changing climate, at least when considered globally.
While our work focuses on 2m temperature, the high forecast skill holds also for other variables.

Currently, all models show a cold bias over land in the forecasts from warmer climate states,\,i.e., the forecasts drift towards the training data compared to the initial conditions. A similar result is obtained for a colder climate state where AIFS and GraphCast show a warming, while Pangu-Weather consistently cools independent of the climatic state. 
GraphCast shows a less pronounced cold bias in the mid-21st century climate, especially for the global mean, where this is, however, achieved by a compensating warm bias over ocean regions. AIFS and GraphCast are overall similar models and show correlated forecast skill in present-day climate (Fig.\,\ref{fig:rmse}b). Therefore, it is currently unclear what causes the observed differences in the future climate. Preliminary experiments indicate that a purely graph-based variant of AIFS (v0.1), which is in principle even closer to GraphCast in terms of methodology (albeit being trained on 1\,$^\circ$ data), shows an even stronger cold bias in 2049 than the current AIFS\,v0.2.1 (not shown).
We believe that other variants of the AIFS that are currently being developed for present-day forecasts, e.g.\,an AIFS with more land and ocean variables or with a different loss normalization will help to shed light on the question of generalization across different climate states.

Future work will explore whether a richer Earth system state, including ocean, land, cryospheric, and external forcing information, can alleviate some of the biases we have observed. 
Consistent cooling patterns over land across all data-driven models point to systematic shortcomings that might be alleviated with the introduction of those additional climate-relevant variables.
As already indicated by our results, 
sea ice cover as additional model input is a promising candidate for an improved performance in changing climate conditions. 
Adding other ocean variables such as mixed-layer depth will be important, as ocean-atmosphere interactions through sea surface temperature (SST)-low cloud feedback can shape atmospheric properties \citep{Norris:1998,Athanase2024projected}.
A similar improvement could be expected from adding land variables with a strong imprint on 2m temperature. The availability of soil moisture does impact the magnitude of current and future continental heatwaves \citep{Miralles2014mega,SanchezBenitez2022nudging} and is likely difficult to be inferred by data-driven models from other parameters.


The present work is a first step towards a thorough comparison and analysis of the  generalization capabilities of different data-driven models. 
It would be important to test the robustness of our results with future states from different climate models, including coarse-resolution CMIP6 results~\citep{Eyring2016cmip6}. As shown recently~\citep{Koldunov2024}, weather forecasting models trained on $0.25^{\circ}$ ERA5 data can perform inherent downscaling of coarser initial conditions. Using data-driven weather forecasting models trained for very high resolution data~\citep{bodnar2024aurora} could provide global information with local granularity for adaptation and mitigation purposes at sub-km scale and at small computational costs. However, the data-driven results will still follow the climate trajectory realised by the coarse-resolution model, as also mentioned by \cite{Koldunov2024}. Therefore, the small scales (e.g. mesoscale ocean eddies, storms, and clouds) and their climate-relevant feedbacks to the large scale will still be overlooked.

One possible other direction is to extend data-driven weather forecasting models to longer integration times, as has already been explored \citep{Wattmeyer2023ace,Guan2024,Wang2024coupled}. Many questions remain open in this context, e.g.\,whether reanalysis or simulation data should be used for training or how the model could be parametrized with information on external forcing so that scenario runs can be generated.
It is currently also unclear how much explicit knowledge about physical processes should be integrated into the data-driven models~\citep{Beucler2024}.
Our work suggests a hybrid approach where conventional climate simulations provide the backbone and data-driven models are used to upsample their output in space, time, or variability. 
This could help, for example, to reduce the storage requirements for climate simulation data, which become prohibitive at km-scales, by using a data-driven weather forecasting model as dynamic interpolation engine to reconstruct in-between states from a sparse set of output fields. 
Another direction would be to use diffusion ensemble AI weather models~\citep{Price2024,AIFSens2024} to estimate the uncertainty around a kilometre-scale climate run, such as those performed in the Climate Change Adaptation Digital Twin developed in the Destination Earth initiative \citep{Hoffmann:2023, Sandu:2024} of the European Commission. However, as our work shows, this will require careful training towards this objective to obtain a well-calibrated ensemble and uncertainty estimates. 
Despite the current limitations highlighted in this study, our results and the fast throughput suggest that data-driven machine learning models will transform our approach towards climate projections and uncertainty quantification in the near future, for example as envisioned in Destination Earth, providing a powerful complement to conventional physics-based models. 

\section*{Methods}

\subsection*{The ai-models package}

The data-driven machine learning models AIFS, GraphCast, and Pangu-Weather are employed using the ai-models package \citep{Raoult2024aimodels}. The Python package has been created by ECMWF to facilitate running this new class of weather forecasting models through a common interface. All models produce 10-day forecasts with 6-hourly time steps in less than 2 minutes.

For the year 2049, all models have been run on ECMWF's ATOS supercomputer. Inputs to the models have been prepared in GRIB format from the nextGEMS IFS-FESOM scenario simulation (see input details below).
For 1955, inputs for ai-models have been retrieved in GRIB format from ERA5 reanalysis for GraphCast and Pangu-Weather, while AIFS has been run via ECMWF's prepml tool (also with ERA5 reanalysis as input). For 2023, daily available forecasts of the three models have been retrieved from ECMWF' experimental suite (\url{https://www.ecmwf.int/en/forecasts/dataset/machine-learning-model-data}). 

\subsection*{Computation of global RMSE and mean bias}

For each 10-day forecast for 2m temperature, denoted as $\mathrm{forecast}(x_i,t)$, root-mean square error is computed as 

$$\mathrm{RMSE}(t) = \sqrt{\frac{1}{\sum w_i} \sum_{i} (\mathrm{forecast}(x_i,t) - \mathrm{reference}(x_i,t))^2 \times w_i},$$

where $t$ is time, $x_i = (lon_i,lat_i)$ is one grid point, $w_i = \cos(lat_i)$ is a latitude weighting that compensates for grid box distortion towards the poles, and $\mathrm{reference}(x_i,t)$ is the reference data from ERA5 (1955), operational analysis (2023), and the reference realisation from the free-running IFS-FESOM simulation (2049).
Similarly, weighted mean bias is computed as
$$\mathrm{mean\,bias}(t) = \frac{1}{\sum w_i} \sum_{i} (\mathrm{forecast}(x_i,t) - \mathrm{reference}(x_i,t)) \times w_i.$$
For the panels in Fig.\,\ref{fig:rmse}, all timeseries in 1955, 2023, and 2049 of $\mathrm{RMSE}_j(t)$ are averaged over all forecasts, i.e. $j=1,...,365$.


\subsection*{AIFS v0.2.1, GraphCast, and Pangu-Weather Models}

The Artificial Intelligence Forecasting System (AIFS) \citep{lang2024aifs} is a data-driven forecast model developed at ECMWF. It uses an encoder-processor-decoder architecture \citep{battaglia2018}. In AIFS v0.2.1, the encoder and decoder are attention-based graph transformers, while the processor is a transformer with a sliding attention window (see Figure 2 in \cite{lang2024aifs}). The AIFS is trained on the ERA5 reanalysis \citep{hersbach2020era5} for the years 1979--2018, and subsequently fine-tuned on ECMWF operational analysis data for 2019--2020. The training objective is a weighted mean squared error, optimized over a forecast horizon of up to 72h; we refer the reader to \cite{lang2024aifs} for more details on the training schedule. The AIFS produces forecasts at the native ERA5 resolution, i.e, on the N320 reduced Gaussian grid at 0.25$^\circ$ resolution.

GraphCast \citep{lam2023graphcast} is a pure graph neural network-based architecture. We used its operational version. It is trained for up to 12 steps, with each having a lead time of 6h, and to provide the time-variance normalized correction to the last time step. The GraphCast network uses an MLP-based graph neural network for encoding from physical space to the network's latent space and also for the decoder that maps back to physical space. The backbone processor of GraphCast uses a 6-level hierarchical mesh based on an octahedral subdivision of the sphere.

PanguWeather \citep{bi2023pangu} is a Swin-Transformer-based architecture and consists of four models for 1h, 6h, 12h and 24h hours lead time. To avoid error accumulation, longer forecasts are achieved by using the combination of models that uses the least steps. No training for multi-step predictions is performed. The Swin-Transformer neural network consists of a U-Net encoder/decoder architecture with patches of size 4x4 in the horizontal dimension. Training used a hand-crafted weighting of different fields with more emphasis on upper air variables.

Like AIFS, Pangu-Weather and Graphcast are trained on 13-pressure levels of standard upper air variables and surface variables of the ERA5-reanalysis from 1979--2018 at the full 0.25$^\circ$ resolution. A comparison of different design choices for machine learning-based medium-range weather forecasting models can be found in~\cite{Nguyen2023_Stormer}.

\subsection*{Initial conditions}

AIFS v0.2.1 takes as initial conditions data on an N320 grid at two successive timesteps. The data includes the 3-dimensional (3D) state variables \texttt{q}, \texttt{t}, \texttt{u}, \texttt{v}, \texttt{w}, and geopotential \texttt{z} on pressure levels, namely at 50\,hPa, 100\,hPa, 150\,hPa, 200\,hPa, 250\,hPa, 300\,hPa, 400\,hPa, 500\,hPa, 600\,hPa, 700\,hPa, 850\,hPa, 925\,hPa, and at the surface (1000\,hPa), see also Table 1 in \citep{lang2024aifs}. Two-dimensional (2D) inputs to AIFS are the 10-metre winds \texttt{10u} and \texttt{10v}, 2-metre (dew point) temperatures \texttt{2d} and \texttt{2t}, mean-sea level pressure (\texttt{msl}), skin temperature (\texttt{skt}), surface pressure (\texttt{sp}), and total-column water (\texttt{tcw}). Four constant fields need to be provided as input as well, namely the land-sea mask (\texttt{lsm}), the surface orography (\texttt{z}), and the standard deviation and slope of the sub-gridscale orography (\texttt{sdor} and \texttt{slor}). 

For the future state, we use initial data from the km-scale  production simulation with IFS-FESOM \citep{rackow2024}, performed in the European Horizon 2020 project nextGEMS (\url{https://nextgems-h2020.eu}), that shows more than 2\,K of warming until the year 2049 compared to the pre-industrial level~\citep{Cycle4:2024}. The simulation uses 9\,km atmospheric resolution and a multi-resolution ocean grid with an average resolution of 5\,km.
The simulation is the first-of-its-kind km-scale climate projection for the period 2020--2050 and has been run on the Levante supercomputer at DKRZ in early 2024.
All 2D and 3D variables have been retrieved directly on the Levante supercomputer, from the nextGEMS simulation, using the \texttt{fdb-read} tool.
Total-column water \texttt{tcw} was computed from its 5 constituents (\texttt{tcwv+tclw+tciw+tcrw+tcsw}) that had been saved in the nextGEMS project. All data have then been remapped to the target N320 grid using the interpolation tool MIR (\url{https://github.com/ecmwf/}).
The four constant fields related to the land-sea mask and orography have been retrieved from ECMWF's operational analysis at N320 resolution.
For every AIFS forecast initialised from nextGEMS data, input variables at times 0600 and 1200 have been combined into a single GRIB1 file.

Pangu-Weather and GraphCast initial conditions for 2049 are taken as a subset from the AIFS ones, and then remapped to regular 0.25\,$^\circ$ grids using MIR. While GraphCast also expects 2 time steps (here 0600 and 1200), Pangu-Weather just expects one (here 1200). 


For the past state, data from the year 1955 from the ERA5 back-extension with anomalously cold global temperatures compared to ERA5 1979-2020 is used as initial condition, retrieved from ECMWF's MARS. We use this year as a proxy for pre-industrial conditions, as reference datasets for weather forecast validation are not available from actual pre-industrial times. Compared to the 1979--2020 period with an average number of 10--20 thousand observations over land per day, about 5 thousand observations over land per day were still available in 1955. 1955 is thus better constrained by data than the period 1940--1950, where occasionally less than 1,000 observations over land are available (personal communication, Bill Bell).
For the present day (2023), the models are initialised from the IFS operational analysis.

\subsection*{Verifying analyses and reference datasets}

It is standard practice to evaluate weather forecasting models against their own analysis. For our 1955 experiment, this is ERA5 
\citep{Bell2021}), which used IFS Cycle 41r2.
For 2023, the IFS analysis is used, which is well-constrained by observations. The analysis for 2023 is based on IFS Cycle 47r3 until 27 June 2023, and Cycle 48r1 after. 
For 2049, the reference dataset is IFS-FESOM (base IFS version 48r1 + modifications that made it into 49r1 \citep{rackow2024}).
Although we acknowledge that biases can be impacted by comparing to different ``truths'', sensitivity tests when exclusively using nextGEMS IFS-FESOM scenario data from different years as input and for reference (e.g. 2020, 2025, 2049), instead of a mix of ERA5, operational analysis, and free-running IFS as reference datasets, gives similar results in terms of RMSE behaviour over 10 days, and a strong cold bias only in future conditions from 2049.
Apart from the physical differences arising from the changing climate states, other technical distinctions (e.g. in terms of IFS cycle) between the reference datasets thus do not seem to be a leading-order effect for the results presented in this study.

\section*{Acknowledgments}

The nextGEMS simulations used in this work and our interactive analysis used supercomputing resources of the German Climate Computing Centre (Deutsches Klimarechenzentrum, DKRZ) granted by its Scientific Steering Committee (WLA) under project ID 1235. This research has been supported by the European Commission Horizon 2020 Framework Programme nextGEMS (grant no. 101003470). This work was also supported by the European Union’s Destination Earth Initiative and relates to tasks entrusted by the European Union to the European Centre for Medium-Range Weather Forecasts implementing part of this Initiative with funding by the European Union. This work is also supported by projects S1: Diagnosis and Metrics in Climate Models of the Collaborative Research Centre TRR 181 “Energy Transfer in Atmosphere and Ocean”, funded by the Deutsche Forschungsgemeinschaft (DFG, German Research Foundation, project no. 274762653). It is also supported by the EERIE project (Grant Agreement No 101081383), funded by the European Union, and by the HClimRep project, funded by the Helmholtz Foundation Model Initiative (HFMI).
We thank Benoit Vanniere and Bill Bell for helpful discussions regarding this study. We thank our colleagues Peter Dueben, Simon Lang, and Zied Ben Bouallegue for providing feedback on this study. We also thank the whole AIFS team for their model development efforts and the ai-models team for their work in facilitating the running of experiments with GraphCast, Pangu-Weather, and the AIFS. Simulations have been performed on the GPU partition of ECMWF's Atos supercomputer.

\bibliographystyle{unsrtnat}

\newpage

\setcounter{figure}{0}
\renewcommand{\thefigure}{S\arabic{figure}}
\section*{Supplementary Material}

\begin{figure}[h]
  \centering
    \includegraphics[width=0.9\textwidth]{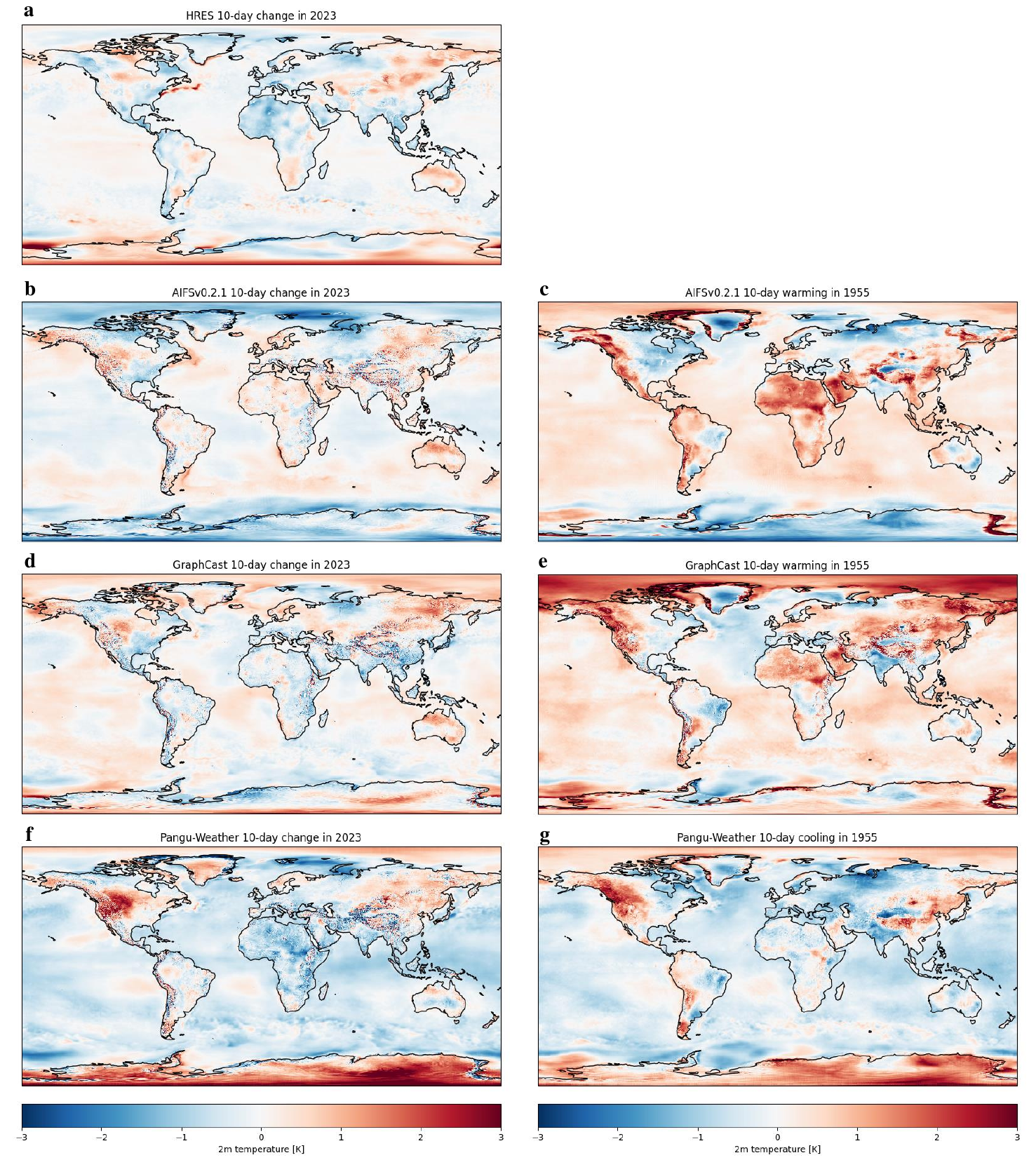}
  \caption{\textbf{Mean 2m temperature drift over 10 days in 2023 (left column) and 1955 (right column) from the data-driven models (and operational IFS forecasts in 2023) when compared to their reference.} References are the operational analysis in 2023 and ERA5 back extension data in 1955. a) Operational IFS forecasts (HRES), b,c) AIFS v0.2.1, d,e) GraphCast, and f,g) Pangu-Weather. Annual-mean fields for the data-driven models were constructed by combining 365 daily means from the end of the individual 10-day forecasts into an annual dataset, and the annual-means from the references were then subtracted.}
  \label{fig:maps1955-2023}
\end{figure}

\end{document}